# Regular phase operator and SU(1,1) coherent states of the harmonic oscillator

**Sándor Varró[1,2]**

1. Wigner Research Centre for Physics of the Hungarian Acadademy of Sciences, Institute for Solid State Physics and Optics, H-1525 Budapest, Pf. 49, Hungary, E-mail: varro.sandor@wigner.mta.hu
2. ELI-ALPS, ELI-Hu Nkft. H-6720 Szeged, Dugonics tér 13, Hungary, E-mail: Sandor.Varro@eli-alps.hu

**Abstract.**
A new solution is proposed to the long–standing problem of describing the quantum phase of a harmonic oscillator. In terms of an 'exponential phase operator', defined by a new 'polar decomposition' of the quantized amplitude of the oscillator, a regular phase operator is constructed in the Hilbert-Fock space as a strongly convergent power series. It is shown that the eigenstates of the new 'exponential phase operator' are SU(1,1) coherent states associated to the Holstein-Primakoff realization. In terms of these eigenstates the diagonal representation of phase densities and a generalized spectral resolution of the regular phase operator are derived, which suit very well to our intuitive pictures on classical phase-related quantities.

## 1. Introduction

The search for a well-defined phase operator (or a quantum angle variable) of a harmonic oscillator has a long history, which began already at the very first years of the foundation of quantum mechanics. In his fundamental paper on the quantum theory of emission and absorption of radiation, Dirac (1927) introduced the quantized amplitudes of the normal modes of the radiation field in 'polar form', which is analogous to the polar decomposition of complex numbers. In the same year, but earlier than Dirac's paper appeared, London (1927) published an analysis on the angle variables and canonical transformations in quantum mechanics, and he proved that the angle variable cannot be expressed by a hermitian matrix. In contrast to the quantum phase (angle), the formal exponential function of it (the 'ladder operators', which also appeared in Dirac's formalism in the 'polar decomposition') could be represented by well-defined matrices. Dirac (1927) quoted London's paper, which clearly shows that he knew about this discrepancy. In spite of this imperfection, the new theory was capable to give a correct description of various fundamental processes taking place in light-matter interaction. As Jordan (1927) noted; „Thus, one is not allowed, for instance, for the action and angle variables, $J$ and $w$, respectively, to write $Jw - wJ = h/2\pi i$ (5). That it was possible to derive several correct results from this not-allowed equation (5), according to a remark of Dirac, is to be understood so that for the derivation of these results, in fact, instead of (5) only $Je^{2\pi i w} - e^{2\pi i w}J = he^{2\pi i w}$ (6) has been used." Here $h$ is Planck's constant, and in the context of the harmonic oscillator $J$ (called „index matrix" by London) corresponds to the number operator $N$, and $w$ to the quantum phase. The formal exponential expression $e^{2\pi i w}$ corresponds to the ladder operator $E^{+}$ (see Section 2). The non-unitarity of $E^{+}$ has also been emphasized by Fock (1932) in the context of second quantization of particle systems. Thus, we may safely say that the founders of the quantum theory were well aware of the fact that Heisenberg's commutation relation $PQ - QP = h / 2\pi i$ for a Cartesian coordinate $Q$ and its canonically conjugate momentum $P$, cannot be taken over for action–angle variables (see also additional notes in the reference to Dirac's quoted paper). It should be noted that, in fact, the first version of field quantization (or in other words, the quantization of a continuous dynamical system) in terms of normal modes have been published by Born, Heisenberg and Jordan (1925) in their famous „Drei Männer Arbeit" (three men's work). They quantized the field amplitudes in term of the Cartesian components $Q$ and $P$, by imposing Heisenberg's commutation relation, thus they did not encounter the problem of quantum action and phase (for further details see Varró (2006)).

In the meantime, there has been much effort devoted to solve the 'problem of the non-unitarity of the ladder operators and of the non-existence of a Hermitian phase operator'. Besides the study of purely mathematical aspects of these problems, the main interest in quantum optics has been to find a well-working formalism for describing quantal phase properties of the electromagnetic radiation, so that these properties could be reliably reconstructed from experimental data. In the frame of the present paper it is not possible even to mention all the important works which have been published on this subject. Still, in the rest of this introduction we shall attempt to briefly summarize the main approaches and results concerning the investigation of the quantum phase problem.

The first period of increasing interest in the nature and description of phase in quantum optics appeared in the sixties of the last century, after the construction of the first lasers. The quantum coherence theory of optical fields has been founded, primarily by Glauber (1963) and by Klauder and Sudarshan (1968) in terms of expectation values of multi-linear expressions of quantized amplitudes, where the action-angle description has no explicit relevance. The non-existence of a Hermitian phase operator of a harmonic oscillator was newly realized by Susskind and Glogower (1964), who seemingly did not know the early discussions by Jordan (1927), London (1927) and Dirac (1927). Susskind and Glogower (1964) introduced *formally* the Hermitian 'cosine' and 'sine' operators, $C = (E + E^{+}) / 2$ and $S = (E - E^{+}) / 2i$, respectively, whose expectation values for a highly excited coherent state approach the cosine and sine of the phase of the complex parameter of such states. Carruters and Nieto (1968) derived various number-phase uncertainty relations by using this 'cosine' and 'sine', and Jackiw (1968) constructed a "critical state" which minimizes one of these uncertainty products. He has also noted that  "unfortunately these states do not seem to have any physical significance".  After a long time it has turned out  that the photon number distribution belonging to Jackiw's minimum-uncertainty state may still be given a quite fundamental physical meaning. Namely, this distribution can also be derived from continuously entangled photon–electron  states from the exact  states of the jointly interacting subsystems (Varró 2008, 2010). Garrison and Wong (1970) constructed a quantum  analogon of the classical periodic phase function (saw-tooth), which satisfies the Heisenberg commutation relation with the number operator on a dense set of the Hardy-Lebesgue space, associated to the oscillator. They have also given an iterative procedure to construct the eigenstates of this phase operator. In our opinion this was the first mathematically correct approach toward the solution of the *original problem* of quantum  phase. Paul (1974) has proposed an alternative description of the phase of a microscopic electromagnetic field, and discussed the possibilities of its measurement.

The second period of interest in the quantum phase problem began around the end of the 1980s, has been intitiated by the growing activity in describing non-classical states, like squeezed states. Pegg and Barnett (1989) has constructed a phase operator by using Loudon's (1973) phase states with equally spaced phase parameters, in a finite-dimensional Hilbert space. We note that the possibility of using a finite-dimensional space in this context has already been discussed by Jordan (1927). The differences of the  Garrison-Wong (1970) and  Pegg-Barnett (1989) formalism have been thoroughly investigated by Popov and Yarunin (1992) and Gantsog *et al* (1992).  The limit matrix elements of the phase operator in number representation (as we let the dimension of the Hilbert space going to infinity) obtained by these authors have already been presented by Weyl (1931). Various 'polar decomposition' of the quantized amplitude have been proposed (Zak 1969, Lerner *et al* 1970, Bergou and Englert 1991) and the non-linear coherent states associated to them have also been investigated for describing real physical processes (de Matos Filho and Vogel 1996, Man'ko *et al* 1997). Recently Urizar-Lenz and Tóth (2010) have  used the old idea of Lévy-Leblond (1976) on the "spectral extension" (unfortunately, later denoted by the same letter as the variance) of non-Hermitian operators for quantifying amplitude squeezing without reference to a phase operator. Adam et al (2014) numerically determined the intelligent states minimizing the 'number-phase uncertainty relation' associated to the number operator and the annihilation operator. The phase distribution of a highly-squeezed states has been determined by Schleich *et al* (1989) by using the radial integral of  the Wigner function (see Schleich 2001) after changing the original (Cartesian) variables to polar coordinates. We note that in the modern era London's work (1927) was quoted for the first time in this paper. Smith *et al* (1992) derived the matrix elements of a phase operator directly, on the basis of a windowed quantum phase-space distribution, and Royer (1996) generalized this approach (see also the book by Dubin *et al* (2000). Lukš and Peřinová (1993) constructed a number-phase Wigner function, depending on a discrete and on a continuous variable. Vaccaro (1995) developed further this approach, and found the satisfactory solution to the problem of fractional photon numbers, which appeared in the former work. Quantum phase measurements has been discussed by Shapiro and Shepard (1991), partly on the basis of "normalizable phase states", which are eigenstates of the usual lowering operator. The completeness integral of these states, however, does not give a meaningful result (see Section 4). No such kind of problem appears when one uses more general SU(1,1) coherent states (see: Vourdas 1990, 1992, 1993; Vourdas and Wünsche 1998, Rasetti 2004). Brif and Ben-Aryeh (1994) have discussed the 'coherent' states introduced by Barut and Girardello (1971), as an example for the 'intelligent states'. Our present study relies, to a considerable extent, on the SU(1,1) coherent states in a Holstein-Primakoff type realization, which was first discussed in this context by Aharonov *et al* (1973) (see also Katriel *et al* 1986, Brif 1995, Gerry and Grobe 1997).

Noh *et al* (1991, 1992a-b, 1993) have investigated the quantum phase dispersion on the basis of *operationally defined* cosine and sine operators, stemming from photon number counts in an eight-port interferometer. Freyberger and Schleich (1993) have quite satisfactorily interpreted the experiment by Noh *et al* (1991). The experimental results of Smithey *et al* (1993) have confirmed the predictions based on both the Pegg-Barnett formalism (1989) and on the radially integrated Wigner distribution, at least for squeezed vacuum states (Schleich *et al* 1989). For further reading and references see the topical issue of *Physica Scripta*, edited by Schleich and Barnett (1993), in which also some historical aspects are summarized by Nieto (1993), the review by Lynch (1995) and the book by Peřinová et al (1999) on the description of phase in optics. We also refer the reader to the review by Dodonov and Man'ko (2003a) in the book on non-classical states in quantum physics (Dodonov and Man'ko, 2003b).

A possible solution to the problem of non-unitarity of the 'exponential phase operator' can be given by a unitary extension ('dilatation') to a larger Hilbert space (see Newton 1980, Stenholm 1993). In fact, the general dilatation theorem was already proved by Sz.-Nagy in 1953 (see the appendix in the book of Riesz and Sz.-Nagy (1965), and references therein). The problem of non-unitarity does not show up in the operational approach developed by Noh *et al* (1991, 1992a-b, 1993), as well as for relative phases of two modes in their product Hilbert space (Luis and Sánchez-Soto 1996), either. Concerning further developments of the mathematical description of the quantum phase of a linear oscillator, see e.g. Kastrup (2003, 2006, 2007) and Rasetti (2004). The first systematic studies of the quantum phase (time-shifts) in a general mathematical framework are due to Helstrom (1974, 1976) and Holevo (1973, 1978, 1979). Concerning recent developments, we refer the reader to the thorough analyses by Lahti and Pellonpää (1999, 2000), Busch *et al* (2001), Pellonpää (2002) and Heinosaari and Pellonpää (2009), in which further development and new results can be found in this direction.

In the present paper a new approach is proposed for treating the quantum phase of the harmonic oscillator. We use the infinite-dimensional Hilbert-Fock space throughout, and will not rely on finite-dimensional subspaces of it. A particular emphasis will be put on the proper formulation and domain of the convergence of operator serieses which define the quantum-phase-related quantities. In this way we avoid ambiguities and inconsistencies concerning, for instance, the exchange of taking expectation values in finite subspaces and going over to the infinite-dimensional Hilbert space. In Section 2 we present the basic definitions concerning the usual 'polar decomposition' and consider some mathematical subtleties of the phase operator introduced by Garrison and Wong (1970). In Section 3 we shall introduce an exponential phase operator based on a new polar decomposition of the quantized amplitude of the harmonic oscillator, in terms of which we define the new 'regular phase operator'. This phase operator is covariant phase observable, according to the definition used by Lahti and Pellonpää (1999, 2000) and Busch *et al* (2001). Section 4 shall be devoted to the introduction of the eigenstates of the new 'exponential phase operator', which will turn out to be SU(1,1) coherent states in Holstein-Primakoff realization. These states form a suitable complete set, in terms of which the phase operator and the corresponding phase densities can be expressed in simple and elegant forms, which correspond to our classical intuition. In Section 5 we shall derive positive operator valued measures for the generalized spectral resolution of the regular phase operator, and introduce the associated probability distribution and density functions. In Section 6 we shall summarize and briefly interpret our results. In order to make our paper possibly self-contained, the details of the mathematical derivations are presented in Appendices A and B.

## 2. The usual 'polar decomposition' of the quantized amplitude, the lack of unitarity of the 'exponential phase operator' and the Garrison-Wong phase operator

The present section will be devoted, on one hand, to the mathematical formulation of the problem we are discussing, and, on the other hand, it is also ment to serve as an explaination of our motivation. We shall present the basic definitions concerning the usual 'polar decomposition' and consider some mathematical subtleties of the phase operator, introduced by Garrison and Wong (1970), and thoroughly discussed also by Popov and Yarunin (1974, 1992). We shall also deal with the question of convergence concerning the series representation of both the classical and the Garrison-Wong phase.

Since the results of the present paper rely on a special, new 'polar decomposition' of the quantized amplitude of a linear oscillator, first we briefly show few details appeared already in the original treatments, mentioned in the introduction. Dirac (1927) considered the amplitudes „$b_r$" and

„$ihb_r^{*}$" to be canonically conjugate variables, and made the contact tranformation, by introducing „$N_r$" as a „probable number of systems in the state $r$", and a new phase „$\theta_r$". The amplitudes are expressed in the 'polar forms' $b_r = e^{-i\theta_r/h} N_r^{1/2} = (N_r+1)^{1/2} e^{-i\theta_r/h}$ and $b_r^{*} = N_r^{1/2} e^{i\theta_r/h} = e^{i\theta_r/h}(N_r+1)^{1/2}$, and they satisfy the *quantum condition* $b_r ihb_r^{*} - ihb_r^{*} b_r = ih$. (In Dirac's notation $r$ refers to a degree of freedom (it labels a normal mode of the radiation field), $h$ is Planck's constant divided by $2\pi$ and $*$ denotes hermitian conjugation. Dirac sometimes explicitly wrote out the symbolic 'representation' $e^{\pm i\theta_r/h} = e^{\mp \delta/\delta N}$ for the 'exponential phase operators', with the help of which the shift relations (see equation (2.5) below) can be 'derived'.) The quantum condition is formally consistent with the commutation relation $\theta_r N_r - N_r \theta_r = ih$ (where, interestingly, the „phase" $\theta_r$ had a dimension of action). Henceforth we shall use dimensionless quantities, and the following notations (for one mode): $b \to A$, $b^{*} \to A^{+}$, thus $[A, A^{+}] \equiv AA^{+} - A^{+}A = 1$.

The usual 'polar decomposition' of the amplitude $A$ is written as

$$A = E\sqrt{N}, \quad A^{+} = \sqrt{N}E^{+}, \quad N = A^{+}A, \tag{2.1}$$

where the 'ladder operator' $E$ is also called 'exponential phase operator', and, occasionally is denoted by $E = \hat{e}^{i\phi}$. The Hilbert space $H$ we are using throughout the present paper is the Fock space spanned by the eigenvectors of the number operator $N$, whose domain $\mathscr{D}(N)$ consists of those vectors for which $N$ has a finite second moment (see e.g. Garrison and Wong 1970 or Busch *et al* 2001)

$$N|n\rangle = n|n\rangle \quad (n = 0,1,2\ldots), \quad \mathscr{D}(N) = \left\{ |\psi\rangle = \sum_{n=0}^{\infty} c_n |n\rangle \in H; \ \sum_{n=0}^{\infty} n^2 |c_n|^2 < \infty \right\}. \tag{2.2}$$

The ladder operator $E$ and its adjoint, $E^{+}$, can be symbolically expressed as

$$E = \sum_{n=0}^{\infty} |n\rangle\langle n+1|, \quad E^{+} = \sum_{n=0}^{\infty} |n+1\rangle\langle n|, \quad EE^{+} = \sum_{n=0}^{\infty} |n\rangle\langle n| = 1, \quad E^{+}E = 1 - |0\rangle\langle 0|, \tag{2.3}$$

where, in the last two equations, we have also displayed their so-called 'half-unitarity property', which was first published by London (1927), and later by Fock (1932) in the context of second quantization in configuration space. In the mathematical terminology an operator like $E^{+}$ is called a *partially isometric operator* (or simply a *partial isometry*), because, though the effect of $E^{+}$ preserves the norm, its adjoint $E$ does not do so (it is a true *contraction operator*). The basic commutators and the shift relations are

$$[E, E^{+}] = EE^{+} - E^{+}E = |0\rangle\langle 0| \equiv P_0, \quad [N,E] = -E, \quad [N,E^{+}] = +E^{+}, \tag{2.4}$$

$$ENE^{+} = N+1, \quad E^{+}NE = N-1+P_0; \quad E^{+}(N+1)E = N. \tag{2.5}$$

We note that the fourth equation in (2.4) shows the corresponding commutation relation we quoted after Jordan (1927) in the Introduction. The *first* and the *third* shift relations shown in (2.5) can *formally* be derived if we write $E = \exp(i\phi)$ and $E^+ = \exp(-i\phi)$, by assuming the existence of an Hermitian $\phi$, which satisfies the commutation relation $[\phi, N] = -i$. Later we shall need the expressions for the effects of the $k$-th power of $E$ and $E^+$ on a generic state $|\psi\rangle = \sum_{n=0}^{\infty} c_n |n\rangle$, i.e.

$$E^k|\psi\rangle = \sum_{r=0}^{\infty} |r\rangle\langle r+k|\sum_{n=0}^{\infty} c_n|n\rangle = \sum_{r,n=0}^{\infty} |r\rangle c_n \delta_{r+k,n}, \quad E^k|\psi\rangle = \sum_{n=0}^{\infty} c_{k+n}|n\rangle \quad (k \geq 0),  \tag{2.6}$$

$$(E^+)^k|\psi\rangle = \sum_{r=0}^{\infty} |r+k\rangle\langle r|\sum_{n=0}^{\infty} c_n|n\rangle = \sum_{n=0}^{\infty} c_n|k+n\rangle \quad (k \geq 0). \tag{2.7}$$

From (2.7) the isometric property of $E^+$ can be easily sown, i.e. $\| (E^+)^k |\psi\rangle \| = \| |\psi\rangle \|$.

Garrison and Wong (1970) introduced a phase operator on the basis of the Hardy class $H^2$ of functions analytic inside the unit disk $D = \{z = re^{i\theta}, |z| < 1\}$ of the complex plane,

$$\langle g|\Phi'_{GW}|f\rangle = \int_{-\pi}^{\pi} \frac{d\theta}{2\pi} g(e^{i\theta})^*[\theta \cdot f(e^{i\theta})], \quad g(e^{i\theta}) = \sum_{n=0}^{\infty} g_n e^{in\theta}, \quad f(e^{i\theta}) = \sum_{n=0}^{\infty} f_n e^{in\theta}. \tag{2.8}$$

The method used by Garrison and Wong (1970) relies on that "the boundary value of a function $f$ is given by a convergent Fourier series containing no Fourier coefficients $f_n$ with negative $n$". A more detailed formulation of the content of this statement is the following. Each function of $H^2$ is also supposed to have a norm $\| f \|_2$ which is determined by the supremum of the integral of $| f(re^{i\theta})|^2$ with respect to the variable $\theta$ in the interval $(-\pi < \theta < \pi)$. The supremum is taken with respect to the radial variable, as a parameter in the range $(0 < r < 1)$. For all such functions the radial limit $\tilde{f}(e^{i\theta}) = \lim_{r \to 1-0} f(z)$ exists almost everywhere (with respect to the Lebesgue measure) on the boundary circle $|z| = 1$, and defines a function $\tilde{f} \in L_+^2$ of $L^2$ of the unit circle, whose Fourier coefficients 'automatically' vanish for negative $n$-s (Hoffman, 1962). The subspace $L_+^2$ of $L^2$ is isomorphic with $H^2$, and has the same metric structure (owing to the same scalar product), that is why $H^2$ is occasionally called 'Hardy-Lebesgue space'. As has been shown by Garrison and Wong (1970), the functions with the property $f(-1) = \sum_{n=0}^{\infty} (-1)^n c_n = 0$ form a dense set $\mathscr{C}$ in $H^2$, and the canonical commutation relations $[\Phi'_{GW}, N]f = if$ is satisfied for these functions. Unfortunately, the *basis states of* $\mathscr{C}$, constructed by Garrison and Wong (1970) *are not orthogonal* to each other (they are, of course, linearly independent). In our view, this non-othogonality is one of the main obstacles which prevents an easy application of these results. This is also the source of that difficulty that the powers of $\Phi'_{GW}$, expressed in the Fock basis, are very complicated (see Popov and Yarunin, 1992). In this context we note that Gantsog, Miranowicz and Tanaś (1992) have performed a thorough analysis of the phase distributions based on the *eigenstates of* $\Phi'_{GW}$, and made a comparison with the predictions based on the formalism introduced by Pegg and Barnett (1989).

In accord with equation (2.8), the matrix elements of $\Phi'_{GW}$ in the Fock basis are

$$\langle n|\Phi'_{GW}|m\rangle = \begin{cases} 0, & n=m \\ i\dfrac{(-1)^{n-m}}{n-m}, & n\neq m \end{cases}. \tag{2.9a}$$

By now we have put a prime in the notation of $\Phi'_{GW}$, reminding us that Garrison and Wong (1970) have represented the number operator by $N=-i\partial/\partial\theta$, in analogy with the $z-$component of the angular momentum operator $L_z$ for motions in the $x-y$ *coordinate plane* ($L_z e^{im\varphi}=me^{im\varphi}, m=0,\pm1,\dots$). However, for an oscillator the rotation in *phase space* is clock-wise (if we associate $q\rightarrow x$ and $p\rightarrow y$, see e.g. Schleich 2001). This corresponds to the symbolic expression $E=\hat{e}^{i\Phi_{GW}}$ with $\Phi_{GW}=-\Phi'_{GW}$ and to the commutation $[N,\Phi_{GW}]=i$, so that the correct shift relations in (2.5) can also be derived. We note that, by changing the base interval from $(-\pi<\theta<\pi)$ to $(0<\theta<2\pi)$, the matrix elements of $\Phi_{GW}$ become

$$\langle r|\Phi_{GW}|s\rangle = \begin{cases} \pi, & r=s \\ \dfrac{i}{s-r}, & r\neq s \end{cases}. \tag{2.9b}$$

By taking (2.6) and (2.7) into account, and introducing an arbitrary reference phase $\varphi_0$ (corresponding to the base interval $(\varphi_0<\theta<\varphi_0+2\pi)$), the Garrison-Wong phase operator can be witten in the form

$$\Phi_{GW}\equiv\varphi_0+\pi+\sum_{k=1}^{\infty}\frac{i}{k}\Big[E^k e^{-ik\varphi_0}-(E^+)^k e^{+ik\varphi_0}\Big]. \tag{2.10}$$

If we take the references phase $\varphi_0=0$ in (2.10), then the matrix elements of $\Phi_{GW}$ are that shown already in (2.9b). The formal operator series can be considered as a quantum analogon of the famous classical Fourier series of the saw-tooth function (see e.g. Gradshtey and Ryzik 1980),

$$\Phi_{cl}(e^{i\varphi})=\pi+\sum_{k=1}^{\infty}\frac{i}{k}\Big[e^{ik\varphi}-e^{-ik\varphi}\Big]=\pi-2\sum_{k=1}^{\infty}\frac{\sin k\varphi}{k}=\pi-2\frac{\pi-\varphi}{2}=\varphi. \tag{2.11}$$

As is known, though the partial sums of this series are bounded, they suffer from the so-called Gibbs phenomenon, and the series is *not absolutely convergent*. This is illustrated in Fig.1 (b) and (c).

FIGURE 1.:

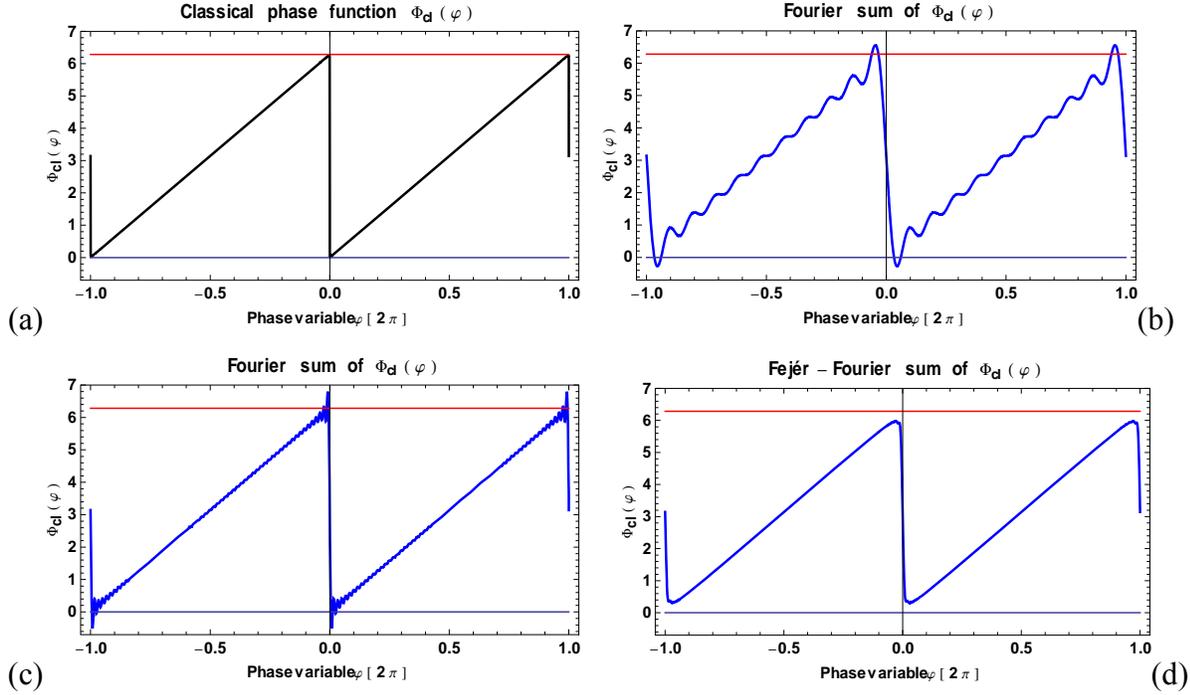

**Figure 1.** Shows (a) the ideal classical phase $\Phi_{cl}(e^{i\varphi})$ as a function of $\varphi$ (measured in $2\pi$ radian units), given by equation (2.11). Partial Fourier sums $s_{11}$ and $s_{61}$ for $n = 11, 61$ are displayed in (b) and (c), respectively. In (d) the $\varphi$-dependence of Fejér mean $\sigma_{61} = \sum_{n=1}^{61} s_n / 61$ is shown. Horizontal lines are also drawn for reference, at the minimum and maximum values, $\Phi_{cl} = 0$ and $\Phi_{cl} = 2\pi$, respectively. The Gibbs phenomenon is clearly seen in (b) and (c) at the discontinuity points $\varphi = -2\pi, 0, 2\pi$. On the other hand, the Fejér mean $\sigma_{61}$ shown in (d) behaves regularly; it stays within the upper an lower bounds (and converges to the arithmetic mean of the left and right limits at the discontinuity points).

It is instructive to compare the usual partial sum $s_n$ and the Fejér mean $\sigma_n \equiv \sum_{m=1}^{n} s_m / n$ (which is the arithmetic mean of the first $n$ partial sums) for the series (2.11), as is shown for $n = 61$ in Figures 1 (c) and (d), respectively. In contrast to the 'singular behaviour' of $s_n$, the Fejér means $\sigma_n$ 'behave regularly', in fact, they *uniformly reconstruct a function* from its Fourier series. Concerning further details on the summability of Fourier serieses see e.g. Titchmarsh (1939), chapter XIII.

On the basis of the conditional convergence of the classical phase function (and the overshoots at the discontinuity points) one would intuitively expect that the convergence of the operator series (2.11) is also 'conditional', in a sense. More precisely, the series of $\Phi_{GW}$ is weakly convergent, though the bilinear form $\langle \chi | \Phi_{GW} | \psi \rangle$ with the kernel martix (2.9b) is bounded for the Hilbert space elements, as has been emphasized by Popov and Yarunin (1974, 1992),

$$\left| \langle \chi | (\Phi_{GW} - \pi) | \psi \rangle \right| = \left| \sum_{n=0}^{\infty} \sum_{m=0}^{\infty} \frac{\langle \chi | m \rangle \langle n | \psi \rangle}{(n-m)} \right| \le \pi \parallel \chi \parallel \cdot \parallel \psi \parallel . \tag{2.12}$$

One should also take into account that this famous inequality (originally derived by Hilbert for *real* sequences) results from the *symmetric* partial double sums

$$T_r = \sum_{n=1}^{r} \sum_{m=1, n \ne m}^{r} \frac{x_n y_m}{n-m} \le \pi \sqrt{\sum_{n=1}^{r} x_n^2} \sqrt{\sum_{m=1}^{r} y_m^2} \tag{2.13}$$

in the $r \to \infty$ limit (see e.g. the book by Cooke (1950), which has also been quoted by Popov and Yarunin (1992) in this context). Accordingly, the two infinite series on the right hand side of (2.10) *separately* does not have a well-defined meaning. Such kind of 'convergence ambiguities' are absent with the 'regular phase operator' to be introduced in the next section. We would also like to point out that in the following considerations we will not rely on any special representation of the Hilbert space of the oscillator (like the Hardy-Lebesgue space, used by Garrison and Wong (1970)).

## 3. Regular phase operator based on a new 'polar decomposition'

In the present section, on the basis of a new 'polar decomposition' of the quantized amplitude $A$, we shall introduce a new 'exponential phase operator' $F$, which depends on a positive parameter $\nu$ (Varró 2014). In terms of powers of $F$ (and of its adjoint $F^+$) we shall define a generalization of the Garrison-Wong phase operator (2.10), which we propose to call 'regular phase operator'.

We define the new 'exponential phase operator' $F$ by the following 'polar decomposition' of the quantized amplitude

$$A = E\sqrt{N} = F\sqrt{N+\nu} \quad (\nu > 0), \quad F = E\sqrt{\frac{N}{N+\nu}}, \quad F^+ = \sqrt{\frac{N}{N+\nu}}E^+, \tag{3.1}$$

where the positive parameter $\nu$ is not subject to any further restrictions. The possible physical meaning of $\nu$ will be discussed in Section 4. Since for both $E$ and $E^+$ are contractions (i.e. $\parallel E \parallel \le 1$ and $\parallel E^+ \parallel \le 1$), and, in general, for any two operators $A$ and $B$ the inequality $\parallel AB \parallel \le \parallel A \parallel \cdot \parallel B \parallel$ holds, we see from (3.1) that $F$ and $F^+$ are contractions, too. With the help of (2.4) and (2.5) one can easily derive the analogous relations,

$$[F, F^+] = \frac{\nu}{(N+1+\nu)(N+\nu)}, \quad [N, F^k] = -kF^k, \quad [N, (F^+)^k] = +k(F^+)^k \quad (k \ge 0), \tag{3.2}$$

$$FNF^+ = (N+1)\left[1 - \frac{\nu}{N+\nu+1}\right], \quad F^+(N+1)F = N\left[1 - \frac{\nu}{N+\nu}\right]. \tag{3.3}$$

From the last two equation of (3.2) it follows that the time-evolution of $F$ and $F^+$ are the same as that of $E$ and $E^+$ (or $A$ and $A^+$), respectively

$$e^{iN\chi}Fe^{-iN\chi} = Fe^{-i\chi}, \quad e^{iN\chi}F^+e^{-iN\chi} = F^+e^{+i\chi}, \tag{3.4}$$

where $\chi$ may be considered as a dynamical phase $\chi = \omega t$ of an oscillator of circular frequency $\omega$. However, according to (3.3), the shift relations (2.5) valid for $E$ and $E^+$ are not preserved for $F$ and $F^+$, but the relation $Fg(N) = g(N+1)F$ holds with an 'arbitrary' function $g(N)$. The explicit expressions for the powers of $F$ and $F^+$ can simply be calculated by using equation (2.3),

$$F^k = f_k^{1/2}(N)E^k, \quad (F^+)^k = (E^+)^k f_k^{1/2}(N) \quad k \geq 0, \tag{3.4}$$

$$f_k(n) \equiv \frac{(n+k)(n+(k-1))\cdots(n+2)(n+1)}{(n+\nu+k)(n+\nu+(k-1))\cdots(n+\nu+2)(n+\nu+1)} = \frac{\Gamma(n+1+k)}{\Gamma(n+1+\nu+k)} \cdot \frac{\Gamma(n+1+\nu)}{\Gamma(n+1)}, \tag{3.5}$$

where $\Gamma$ denotes the usual gamma function (see e.g. Gradshteyn and Ryzhik 1980).

Notice that for any $\nu > 0$, $f_0(n) = 1$ for all $n \geq 0$, and in case of $\nu = 0$, we have $f_k(n) = 1$ for all $n \geq 0$ and $k \geq 0$. Below we shall need certain inequalities satified by the functions $f_k(n)$, defined in (3.5). As is shown in Appendix A (see (A.8)), for any values $\nu > 0$, the following inequalities are valid

$$f_k(0) < \Gamma(1+\nu)e^{\nu+1/6}\frac{1}{(k+\nu)^\nu} \ (k \geq 1, n = 0), \quad f_k(n) < e^{1/4}\sqrt{1+\nu}\frac{(n+\nu)^\nu}{(k+\nu)^\nu} \ (k \geq 1, n \geq 1), \ (\nu > 0), \tag{3.6}$$

By prescribing the correspondence with the classical ('saw-tooth') phase function in (2.11), i.e. by using the associations of $e^{i\varphi} \to F$ and $e^{-i\varphi} \to F^+$, we define the new *regular phase operator* as the infinite series

$$\Phi = \varphi_0 + \pi + \sum_{k=1}^{\infty}\frac{i}{k}\left[F^k e^{-ik\varphi_0} - (F^+)^k e^{+ik\varphi_0}\right]. \tag{3.7}$$

Of course, one has to define the domain of this $\Phi$, too, by considering the convergence properties of the operator series in (3.7). Our aim here is to find a domain in Hilbert space where the right hand side of (3.7) can be represented by a sum of two infinite serieses, which converge strongly and separately. On the basis of the inequalities (3.6), it is shown in Appendix A (see the derivation of (A.14a-b)), for any generic state $|\psi\rangle = \sum_{n=0}^{\infty}c_n|n\rangle \in H$, the upper bounds of the norms of $F^k|\psi\rangle$ and $(F^+)^k|\psi\rangle$ are subject to the uniform estimates

$$\left\| F^k |\psi\rangle \right\|^2 < \frac{\overline{b}_{\nu,\psi}}{(k+\nu)^\nu}, \qquad \left\| (F^+)^k |\psi\rangle \right\|^2 < \frac{\overline{b}_{\nu,\psi}}{(k+\nu)^\nu} \qquad (\nu>0, k\geq 1),$$ (3.8a)

$$\overline{b}_{\nu,\psi} \equiv e^{\nu+1/6}\Gamma(1+\nu) + e^{1/4}\sqrt{1+\nu}\left\langle (N+\nu)^\nu \right\rangle_\psi, \qquad \left\langle (N+\nu)^\nu \right\rangle_\psi = \sum_{n=0}^{\infty}(n+\nu)^\nu \mid c_n \mid^2 < \infty.$$ (3.8b)

It is of crucial importance here that the bound parameter $\overline{b}_{\nu,\psi}$ introduced in (3.8b) does not depend on power index $k$ of $F^k$ (and of $(F^+)^k$, either). As is seen in (3.8b), the bounds in (3.8a) exist and non-trivial only for those states for which the $\nu - th$ *moment of* $(N+\nu)$ *is finite*. We denote the set of such states by $\mathscr{D}_\nu(N)$,

$$\mathscr{D}_\nu(N) = \left\{ |\psi\rangle = \sum_{n=0}^{\infty}c_n|n\rangle \in H; \; \sum_{n=0}^{\infty}(n+\nu)^\nu \mid c_n \mid^2 < \infty \right\}.$$ (3.9)

It is clear that for $\nu -$ values in the range $0 < \nu \leq 2$, the domain $\mathscr{D}(N)$ (see (2.2)) is contained by $\mathscr{D}_\nu(N)$, i.e. $\mathscr{D}(N) \subseteq \mathscr{D}_\nu(N)$, and if $\nu \geq 2$, then $\mathscr{D}_\nu(N) \subseteq \mathscr{D}(N)$. On the basis of these considerations, one finds that the infinite series expressions in the defining equation (3.7) of the new phase operator converge strongly in $\mathscr{D}_\nu(N)$, because, for instance, for the first series we have

$$\left\| \sum_{k=1}^{\infty} \frac{ie^{-ik\varphi_0}}{k} F^k |\psi\rangle \right\| \leq \sum_{k=1}^{\infty}\frac{1}{k}\left\| F^k |\psi\rangle \right\| \leq \sum_{k=1}^{\infty}\frac{(\overline{b}_{\nu,\psi})^{1/2}}{k(k+\nu)^{\nu/2}} \leq (\overline{b}_{\nu,\psi})^{1/2}\sum_{k=1}^{\infty}\frac{1}{k^{1+\nu/2}} = (\overline{b}_{\nu,\psi})^{1/2}\zeta(1+\nu/2),$$ (3.10)

where we have introduced the Riemann zeta function $\zeta(s)$ (see e.g. Titchmarsh, 1939). Owing to the inequalities in (3.8a), the same bound exists for the second series on the right hand side of (3.7). Thus, the two terms of the infinite series in (3.7) *converge separately and strongly* in $\mathscr{D}_\nu(N)$ for any $\nu > 0$, which means, at the same time, that we are justified to write (3.7) in the alternative form

$$\Phi = \varphi_0 + \pi + \sum_{k=1}^{\infty}\frac{i}{k}F^k e^{-ik\varphi_0} - \sum_{k=1}^{\infty}\frac{i}{k}(F^+)^k e^{+ik\varphi_0} = \varphi_0 + \pi - i[\log(1-Fe^{-i\varphi_0}) - \log(1-F^+ e^{+i\varphi_0})],$$ (3.11)

where we have *formally* used the sum of the power series $\sum_{k=1}^{\infty}(x^k/k) = -\log(1-x)$, which is convergent in the interval $(-1 \leq x < 1)$. For $|z| < 1$ the latter numerical series is absolutely convergent, but, of course, for a complex $z$ we have to specify the branch of the logarithm, too. In all the cases where we shall use (3.11) below, the main branch of the logarithm will be taken in the expectation values. The second equality in (3.11) is to be rather considered as the *definition* of the log functions of the operators, which is surely meaningful, at least in the sense of strong convergence in the domain $\mathscr{D}_\nu(N)$. A completely different situation appears if $\nu = 0$ (i.e. $F = E$ and $F^+ = E^+$, in which case the series corresponds to the Garrison-Wong phase operator (2.10)). In this case on the right hand side of (3.10) we encounter with a logarithmically divergent factor $\sum_{k=1}^{\infty}(1/k)$, and the above estimates cannot be used. The limit of the symmetric double sum in (2.13) still exists, and represents the phase operator $\Phi_{GW}$ of Garrison and Wong (1970), as has been emphasized by Popov and Yarunin (1992).

On the basis of the last two equation in (3.2), the 'number-phase commutator' can be brought to the form

$$[N,\Phi] = i - 2\pi i P_{\varphi_0}\,, \quad P_{\varphi_0} \equiv \frac{1}{2\pi}\sum_{k=-\infty}^{\infty} F_k e^{-ik\varphi_0}\,, \quad F_k \equiv \begin{cases} F^k\,, & k \geq 0 \\ (F^+)^{|k|}\,, & k \leq 0 \end{cases}, \tag{3.12}$$

where we have introduced the factor $2\pi$ in front of $P_{\varphi_0}$ for later convenience. From (3.8a) and (3.9) we see that the series in (3.12) converges strongly in $\mathscr{D}_\nu(N)$, if $\nu > 2$, in which case $\mathscr{D}_\nu(N) \subseteq \mathscr{D}(N)$. At the end of the present section we present the matrix elements of $\Phi$ and $P_{\varphi_0}$ in the Fock basis,

$$\langle r|\Phi|s\rangle = \begin{cases} \varphi_0 + \pi\,, & r = s \\ \dfrac{i}{s-r} f_{|r-s|}^{1/2}[\min(r,s)]e^{i(r-s)\varphi_0}\,, & r \neq s \end{cases}, \tag{3.13a}$$

$$2\pi\langle r|P_{\varphi_0}|s\rangle = f_{|r-s|}^{1/2}[\min(r,s)]e^{i(r-s)\varphi_0}\,, \quad 2\pi P_{\varphi_0} = \sum_{r=0}^{\infty}\sum_{s=0}^{\infty}(e^{ir\varphi_0}|r\rangle)f_{|r-s|}^{1/2}[\min(r,s)](\langle s|e^{-is\varphi_0})\,, \tag{3.13b}$$

where the function $f_k(n)$ has been defined in (3.5). We note that Smith *et al* (1992) derived the matrix elements of a phase operator, directly on the basis of the Weyl-Wigner correspondence, but their result is different from (3.13a), as is expected. Since for $\nu = 0$ we have $f_k(n) = 1$ for all $n \geq 0$ and $k \geq 0$ (see the defining equation (3.5) of $f_k(n)$), with the help of the second equation of (3.13b), it can be immediately shown that the operator $P_{\varphi_0}$ can *formally* be expressed as a dyad of the (singular, i.e. *not* normalizable) phase state $|\varphi_0\rangle$,

$$\nu = 0: \ 2\pi P_{\varphi_0} = |\varphi_0\rangle\langle\varphi_0|\,, \quad |\varphi_0\rangle = \sum_{n=0}^{\infty} e^{in\varphi_0}|n\rangle\,. \tag{3.14}$$

According to the considerations after equation (2.8) in the previous section, the states $|f\rangle$ for which $\langle\varphi_0|f\rangle = 0$, just belong to the dense set $\mathscr{C}$ in $H^2$ for which the canonical commutation relation holds (notice that Garrison and Wong (1970) have taken $\varphi_0 = -\pi$ as a reference phase). To prove at least the existence of a similar set of states for our general dyadic expression $P_{\varphi_0}$ in (3.13b) is an open question.

Anyway, according to (3.13b) and (3.14), $\langle\psi|2\pi P_{\varphi_0+\varphi}|\psi\rangle$ can be considered as a generalizaton of the Loudon phase distribution $|\langle\varphi_0+\varphi|\psi\rangle|^2$, which has been widely used. We note that the matrix elements $\langle r|P_{\varphi_0+\varphi}|s\rangle$ have the properties (b) and (c) in the „phase theorem" of Lahti and Pellonpää (1999).

## 4. Regular phase states and the diagonal representation of the regular phase operator

Our formulae for both the number-phase commutator (3.12) and for the matrix elements (3.13a-b) are quite complicated in comparison with those appearing in the Garrison-Wong formalism. In the present section we shall introduce a suitable complete set of states, in terms of which all mathematical objects will receive a simple and elegant form, corresponding to our classical intuition. This correspondence relies on the strong convergence of the defining serieses of our regular phase operator (expressed by (3.11) of Section 3), i.e. we need not care convergence questions concerning the rearrangement of infinite (operator) serieses. We shall prove that the eigenstates of $F$, introduced in (3.1) of the previous section, are SU(1,1) coherent states with Bargmann index $\kappa=\frac{1}{2}(\nu+1)>\frac{1}{2}$, associated to the su(1,1) Lie algebra of a type of Holstein-Primakoff realization *in the original Hilbert space of one harmonic oscillator*. In the last part of the present section we will show that these states serve as a natural basis for representing the new phase operator and other quantum-phase-related quantities.

The normalized solution of the eigenvalue equation $F|z\rangle=z|z\rangle$ (see Appendix B) reads

$$F|z\rangle=z|z\rangle\,;\quad |z\rangle=(1-|z|^2)^{\frac{1}{2}(\nu+1)}\sum_{n=0}^{\infty}\left[\frac{\Gamma(\nu+1+n)}{\Gamma(\nu+1)(n!)}\right]^{1/2}z^n|n\rangle\,,\quad z\in D=\{z,|z|<1\}\,, \tag{4.1}$$

where $z$ is a complex number in the open unit disc $D$. The expansion coefficients of $|z\rangle$ have the same form as that of the SU(1,1) coherent states in the discrete series representation (Perelomov 1986). As is shown in Appendix B, these states represent a scaled variant of the states $|\alpha,k\rangle$ derived by Aharonov *et al* (1973), as eigenstates of a generalized annihilation operator (see equations (B.3d-e) in Appendix B). The expectation value, variance and the Mandel Q parameter of the (photon) number distribution $\{p_n\equiv|w_n|^2,\ n=0,1,...\}$ of an SU(1,1) state $|\zeta=\rho\exp(i\theta)\rangle=\sum_{n=0}^{\infty}w_n|n\rangle$ (see (B.2c)) can be calculated by using the derivatives of the negative binomial power series (B.1c),

$$\langle N\rangle_\zeta=(1+\nu)\frac{\rho^2}{1-\rho^2}\,,\quad \Delta N_\zeta^2\equiv\langle N^2\rangle_\zeta-\langle N\rangle_\zeta^2=(1+\nu)\frac{\rho^2}{(1-\rho^2)^2}\,,\quad Q\equiv\frac{\Delta N_\zeta^2}{\langle N\rangle_\zeta}-1=\frac{\rho^2}{1-\rho^2}>0\,. \tag{4.2}$$

We note that Gilles and Knight (1992) have presented an extensive analysis of the statistical properties of the photon number distribution of the type $\{p_n\equiv|w_n|^2,\ n=0,1,...\}$ of (B.2c), in the context of *two-mode* squeezed states. In this case the physical meaning of the parameter $\nu$ is the invariant difference of the photon numbers, $\hat{n}_a - \hat{n}_b$ of the two modes, whose interaction is represented by the parametric Hamiltonian $\propto i(\lambda \hat{a}\hat{b} - \lambda^* \hat{a}^+ \hat{b}^+)$. Since the *regular phase states are one-mode pure states*, in spite of common features in the photon number distributions with that of the two-mode squeezed states, we cannot give the parameter $\nu$ an immediate meaning on this basis.

The eigenstates in (4.1) are in fact *one-mode SU(1,1) coherent states in Holstein-Primakoff realization* (Aharonov *et al* 1973, Katriel *et al* 1986, Brif and Ben-Aryeh 1994, Brif 1995, Gerry and Grobe 1997), associated to the Lie algebra of the basis operators $K_\pm$ and $K_0$ (see Appendix B),

$$K_- = A\sqrt{N+\nu} = F(N+\nu), \quad K_+ = \sqrt{N+\nu}A^+ = (N+\nu)F^+, \quad K_0 = N + \tfrac{1}{2}(\nu+1) = N + \kappa, \quad (4.3a)$$

$$[K_-, K_+] = 2(K_0 + \kappa), \quad [K_0, K_\pm] = \pm K_\pm, \quad (4.3b)$$

where we have introduced the Bargmann index $\kappa$, in terms of which $\nu = 2\kappa - 1$. It can also be shown that

$$[F, K_0] = F, \quad [F, K_+] = 1, \quad [F, K_-] = F^2, \quad (4.3c)$$

from which it follows that the eigenstates of $F$ are also eigenstates of the commutators (4.3c). On the basis of this property, Aharonov *et al* (1973) have shown that the states $|z\rangle$ are generated from the vacuum state by the unitary transformation,

$$|z\rangle = \exp(\xi e^{+i\theta} K_+ - \xi e^{-i\theta} K_-)|0\rangle, \quad z = \tanh(\xi)e^{i\theta}. \quad (4.4)$$

The interaction terms $\propto K_\pm$ represent a kind of 'intensity-dependent coupling'. We note that the coherent states introduced by Barut and Girardello (1971) are eigenstates of $K_- = F(N+\nu)$, so they differ from $|z\rangle$, of course.

The states $|z = \rho \exp(i\theta)\rangle$ with different $z-$s are not orthogonal, but they form an overcomplete set in the original Hilbert space $H$. The completeness can be shown by introducing a suitable weight function, and then integrating with respect to $\theta$ and $\rho$ over the open unit disc $D$ (see Appendix B),

$$\frac{\nu}{\pi}\int_D d^2\mu(z)|z\rangle\langle z| = 1, \quad d^2\mu(z) \equiv d^2 z(1-|z|^2)^{-2}, \quad d^2 z \equiv d(\mathrm{Re}\,z)d(\mathrm{Im}\,z) = \rho d\rho d\theta \quad (\nu > 0). \quad (4.5)$$

We note at this point that the so-called "normalizable phase states" $|z\rangle_0$, first discussed by Lerner *et al* (1970) and later used e.g. by Shapiro and Shepard (1991) and D'Ariano *et al* (1996), represent a special case of $|z\rangle$ with $\nu = 0$. They are eigenstates of the usual exponential phase operator $E$,

$$E|z\rangle_0 = z|z\rangle_0; \quad |z\rangle_0 = \sqrt{1-|z|^2}\sum_{n=0}^{\infty} z^n |n\rangle, \quad z \in D = \{z, |z| < 1\}. \quad (4.6)$$

Unfortunately, the completeness integral (4.4) does not give a reasonable result in case of $\nu = 2\kappa - 1 = 0$ (see Appendix B), thus one cannot use $|z\rangle_0$ as basis states. In fact, this circumstance has been one of the roots of our motivations for introducing $F$ and its regular eigenstates $|z\rangle$.

Now we are in the position to express the regular phase operator (3.7) in a diagonal form, in terms of the *regular phase states* (4.1). Owing to equation (3.11) and the eigenvalue equation $F|z\rangle = z|z\rangle$ (and also $\langle z|F^+ = z^*\langle z|$), by using the completeness relation (4.5) we receive (see Appendix B)

$$\Phi = \frac{\nu}{\pi}\int_D d^2\mu(z)|z\rangle\varphi(z)\langle z|, \qquad e^{i\varphi(z)} = e^{i(\varphi_0 + \pi)}\frac{1 - ze^{-i\varphi_0}}{1 - z^*e^{+i\varphi_0}}, \qquad (4.7)$$

$$\varphi(z) \equiv \varphi(z,z^*) = \varphi_0 + \pi - 2\arctan\left[\frac{\rho\sin(\theta - \varphi_0)}{1 - \rho\cos(\theta - \varphi_0)}\right] \qquad (\varphi_0 < \varphi(z) < \varphi_0 + 2\pi), \qquad (4.7a)$$

where we have used the parametrization $z = \rho e^{i\theta}$, and introduced the *quantum phase function* $\varphi(z)$. It is important to note that the quantum phase function does not depend on the parameter $\nu$. From the original form (3.7) of the regular phase operator one can show that the diagonal matrix elements of $\Phi$ between the regular phase states just coincide with the quantum phase function $\varphi(z)$ defined in equation (4.7a),

$$\langle z|\Phi|z\rangle = \varphi(z) = \varphi_0 + \pi - 2\sum_{k=1}^{\infty}\rho^k\frac{\sin[k(\theta - \varphi_0)]}{k}. \qquad (4.8)$$

The two forms of $\varphi(z)$ in (4.7a) and (4.8) are equivalent, which can be shown by using the Fourier series of the *arctan* function (see Appendix B). Since the Fourier series $\sum_{k=1}^{\infty}(\sin k\varphi)/k$ converges to $\frac{1}{2}(\pi - \varphi)$ in the interval $(0 < \varphi < 2\pi)$, in the limit $\rho \to 1-0$ the quantum phase function $\varphi(z)$ approaches the classical phase function $\Phi_{cl}(e^{i\varphi})$, given by (2.11) (where we have taken the reference phase $\varphi_0 = 0$). This behaviour is illustrated in Fig.2a.

FIGURE 2.:

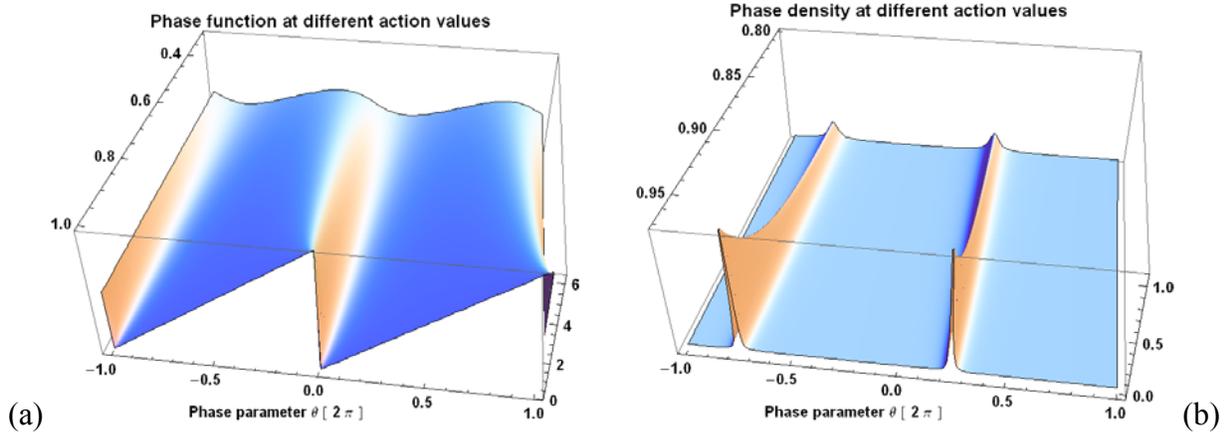

**Figure 2.** (a) Shows the quantum phase function $\varphi(z)$, defined in (4.7a) (and coinciding with (4.8)) for different radial parameter ('action') values $0.3 \leq \rho \leq 0.999$, as a function of the phase parameter in the range $-2\pi \leq \theta \leq +2\pi$. It approaches the classical phase function $\Phi_{cl}(e^{i\varphi})$, given by (2.11) (we have taken the reference phase $\varphi_0 = 0$). (b) Shows the quantum phase density $p_{\varphi_0+\varphi'}(z)$ (which is the classical Poisson kernel, at the same time; see equation (5.1a) in Section 5), as a function of $\theta$, for radial parameter ('action') values $0.8 \leq \rho \leq 0.97$. We have choosen $\varphi' = \pi/2$. As $\rho$ increases the quantum phase density is more and more peaked around $\varphi'$.

## 5. Generalized spectral resolution of the regular phase operator and phase probability distributions

The regular phase operator $\Phi$ defined by (3.7) can also be represented as a one-dimensional integral on the base interval $(\varphi_0 < \varphi = \varphi_0 + \varphi' \leq \varphi_0 + 2\pi)$ of the set of positive operators $P_{\varphi_0+\varphi'}$,

$$\Phi = \int_{\varphi_0}^{\varphi_0+2\pi} d\varphi \, \varphi P_\varphi \, , \qquad P_{\varphi_0+\varphi'} = \frac{\nu}{\pi} \int_D d^2\mu(z) |z\rangle p_{\varphi_0+\varphi'}(z) \langle z| \, , \tag{5.1}$$

$$p_{\varphi_0+\varphi'}(z) = \frac{1}{2\pi} \frac{1-\rho^2}{1-2\rho\cos(\theta-\varphi_0-\varphi')+\rho^2} \, , \tag{5.1a}$$

where the *quantum phase density* $p_{\varphi_0+\varphi'}(z)$ is just the well-known Poisson kernel. We have derived it, by using the formula 1.447.3 of Gradshteyn and Ryzhik (1980). In fact, we also have

$$p_{\varphi_0+\varphi'}(z) = \langle z|P_{\varphi_0+\varphi'}|z\rangle = \frac{1}{2\pi}\left\{1 + 2\sum_{k=1}^{\infty} \rho^k \cos[k(\theta-\varphi_0-\varphi')]\right\} \quad (0 < \varphi' \leq 2\pi) \, . \tag{5.1b}$$

Like the quantum phase function, the quantum phase density does not depend on the parameter $\nu$, either. As a function of $\theta-\varphi_0$, in the limit $\rho \to 1-0$ the Poisson kernel is more and more peaked around $\varphi'$, as is shown in Fig.2b. The explicit form of the $P_{\varphi_0+\varphi'}$, in terms of the operators $F$ and $F^+$ can simply be obtained by applying the diagonal representation of $\Phi$ and the completeness relation, given by (4.7) and (4.5), respectively,

$$P_{\varphi_0+\varphi} = \frac{1}{2\pi}\left\{1 + \sum_{k=1}^{\infty}[F^k e^{-ik(\varphi_0+\varphi)} + (F^+)^k e^{+ik(\varphi_0+\varphi)}]\right\} \equiv \frac{1}{2\pi}\sum_{k=-\infty}^{\infty}F_k e^{-ik(\varphi_0+\varphi)} \quad (0 < \varphi \le 2\pi). \quad (5.2)$$

It is remarkable that we have already encountered with this positive operator in equation (3.12), where we have calculated the 'number-phase commutator' $[N,\Phi] = i - 2\pi i P_{\Phi_0}$.

In Section 3 we have defined the regular phase operator (3.7) by using the correspondences $e^{i\varphi} \to F$ and $e^{-i\varphi} \to F^+$ in the Fourier series of the classical (periodic, 'saw-tooth') phase function (2.11), which converges ('reconstructs') $\varphi$ on the base interval. Similarly, we can associate a set of operators to the periodic functions $e_\psi(\varphi)$ (taking on values 1 or 0) defined as

$$e_\psi(\varphi) = \begin{cases} 1 & if & 2k\pi < \varphi \le 2k\pi + \psi \\ 0 & if & 2k\pi + \psi < \varphi \le 2(k+1)\pi \end{cases} \quad (k = 0, \pm 1, \pm 2\ldots), \quad (5.3)$$

with $e_0(\varphi) \equiv 0$ and $e_{2\pi}(\varphi) \equiv 1$ (Riesz and Sz.-Nagy 1965). The function defined in (5.3), as a function of $\psi$ is similar to a degenerate distribution function, in the terminology of probability theory. In the interval $(0 < \varphi \le 2\pi)$, for instance, it cuts a part $(0 < \varphi \le \psi)$ where it equals to unity, and on the rest of the interval it is zero. This function always equals to its square, i.e. $[e_\psi(\varphi)]^2 = e_\psi(\varphi)$, and, moreover, $e_\chi(\varphi)e_\psi(\varphi) = e_\chi(\varphi)$ for $\chi \le \psi$. Owing to these properties, we shall call $e_\psi(\varphi)$ *classical projector function*, whose plot is shown in Fig. 3a. By the correspondence $e^{i\varphi} \to F$ (and $e^{-i\varphi} \to F^+$) we associate a set of operators $E_\psi$ to the Fourier series representation of $e_\psi(\varphi)$

$$e_\psi(\varphi) = e_\psi(e^{i\varphi}) = \frac{1}{2\pi}\sum_{k=-\infty}^{\infty}\frac{i}{k}(e^{-ik(\psi-\varphi_0)}-1)e^{ik(\varphi-\varphi_0)} \to$$

$$E_\psi = \frac{1}{2\pi}\sum_{k=-\infty}^{\infty}\frac{i}{k}(e^{-ik(\psi-\varphi_0)}-1)F_k e^{-ik\varphi_0} \quad (\varphi_0 < \psi < \varphi_0 + 2\pi), \quad (\varphi_0 < \varphi \le \varphi_0 + 2\pi), \quad (5.4)$$

where we have introduced a reference phase $\varphi_0$, and used the notation $F_k = F^k$ $(k \ge 0)$ and $F_k = (F^+)^{|k|}$ $(k < 0)$, as has been already defined in (3.12). The operator series (5.4) converges strongly in $\mathscr{D}_\nu(N)$. The diagonal expansion of $E_\psi$ can be obtained, by using the completeness relation (4.4) and the eigenvalue equations $F|z\rangle = z|z\rangle$ (and $\langle z|F^+ = z^*\langle z|$),

$$E_\psi = \frac{\nu}{\pi} \int_D d^2\mu(z) |z\rangle e_\psi(z) \langle z|, \tag{5.4a}$$

$$e_\psi(z) \equiv \frac{1}{2\pi}\left\{ \psi - \varphi_0 + \sum_{k=1}^{\infty} \frac{i}{k}\left[ (e^{-ik(\psi-\varphi_0)}-1)(ze^{-i\varphi_0})^k \right] - \sum_{k=1}^{\infty} \frac{i}{k}\left[ (e^{+ik(\psi-\varphi_0)}-1)(z^* e^{+i\varphi_0})^k \right] \right\}, \tag{5.4b}$$

where we have introduced the *quantum projector function* $e_\psi(z)$, which, similarly to $\varphi(z)$ and $p_\varphi(z)$, does not depend on the parameter $\nu$, either. In Figure 3 we have compared the classical and quantum projector functions, $e_\psi(\theta)$ and $e_\psi(z = \rho e^{i\theta})$, defined in (5.3) and (5.4b), respectively.

FIGURE 3 (a).:

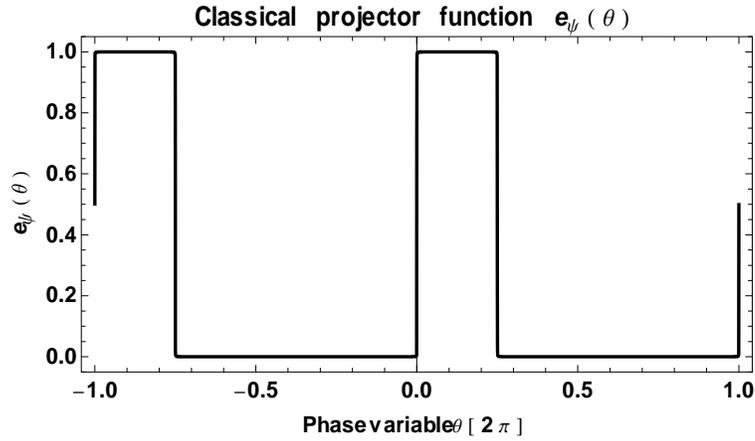

FIGURE 3 (b).:

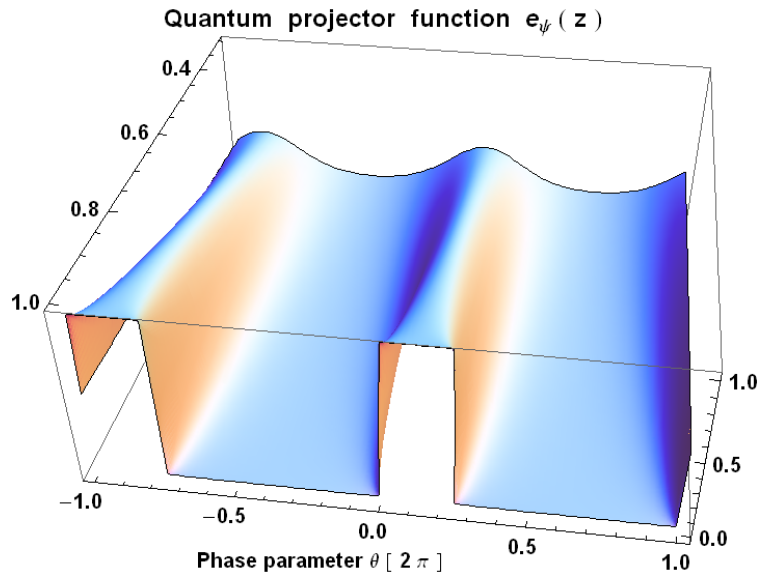

**Figure 3.** (a) Shows the classical projector function $e_\psi(\theta)$, defined in (5.3) as a function of the phase parameter in the range $-2\pi \leq \theta \leq +2\pi$, (b) Shows the quantum projector function $e_\psi(z = \rho e^{i\theta})$, defined in (5.4b), with different radial parameter values $0.3 \leq \rho \leq 0.999$. We have taken the reference phase $\varphi_0 = 0$, and $\psi = \pi/2$.

If $F$ were unitary, then $\{E_\psi\}$ would form a true spectral set (Riesz and Sz.-Nagy 1965) consisting of mutually orthogonal projectors, but this is not the case here. This is just the original problem with the 'exponential phase operators' (with both $E$ and $F$). Still, one can proceed further and derive a 'generalized spectral resolution' of $\Phi$ (which is essentially equivalent with (5.1)). From (5.2) and (5.4) it can be seen that $P_\psi$ is the derivative of $E_\psi$ with respect to $\psi$, thus the one-dimensional integral representation of the phase operator (see the first equation in (5.1)) can also be written as

$$\Phi = \int_{\varphi_0}^{\varphi_0+2\pi} \psi P_\psi d\psi = \int_{\varphi_0}^{\varphi_0+2\pi} \psi \frac{dE_\psi}{d\psi} d\psi = \int_{\varphi_0}^{\varphi_0+2\pi} \psi dE_\psi, \qquad F_k = \int_{\varphi_0}^{\varphi_0+2\pi} e^{ik\varphi} P_\varphi d\varphi = \int_{\varphi_0}^{\varphi_0+2\pi} e^{ik\varphi} dE_\varphi. \tag{5.5}$$

The second set of equations (obtained from (5.2) and $P_\varphi = dE_\varphi/d\varphi$) expresses the *spectral resolution in a wider sense* of the contraction operators $F_k = F^k$ ($k \geq 0$) and $F_k = (F^+)^{|k|}$ ($k < 0$). The last equation is an analogon of the usual spectral resolution of unitary operators (see Riesz and Sz.-Nagy 1965). According to (5.5) the expectation value of the phase operator in a state represented by a density operator $\hat{\rho}$ can be brought to the following equivalent forms,

$$\langle \Phi \rangle = Tr(\hat{\rho}\Phi) = \int_{\varphi_0}^{\varphi_0+2\pi} \psi d[Tr(\hat{\rho}E_\psi)] = \int_{\varphi_0}^{\varphi_0+2\pi} \psi \frac{d\{Tr[\hat{\rho}E_\psi]\}}{d\psi} d\psi = \int_{\varphi_0}^{\varphi_0+2\pi} \varphi Tr[\hat{\rho}P_\varphi] d\varphi. \tag{5.6}$$

Equation (5.6) shows that $Tr[\hat{\rho}E_\varphi]$ can be considered as a *phase probability distribution*, $G(\varphi)$,

$$G(\varphi) \equiv Tr[\hat{\rho}E_\varphi] = \frac{\nu}{\pi} \int_D d^2\mu(z) e_\varphi(z) \langle z|\hat{\rho}|z\rangle, \tag{5.7}$$

$$G(\varphi) = Tr[\hat{\rho}E_\varphi] = \frac{\nu}{2\pi} \int_0^1 d(r^2) \int_0^{2\pi} d\theta \frac{e_\varphi(z)}{(1-r^2)^2} \langle z|\hat{\rho}|z\rangle, \tag{5.7a}$$

where $e_\varphi(z)$ has been defined in (5.4b). The derivative $dTr[\hat{\rho}E_\varphi]/d\varphi = Tr[\hat{\rho}P_\varphi]$ (if it exists) defines the *phase probability density distribution*, $g(\varphi)$. Owing to the second and third equations of (5.1), the phase probability density distribution $g(\varphi)$ is an integral of the product of the 'Q-function' $\langle z|\hat{\rho}|z\rangle$ and the Poisson kernel ($z = r\exp(i\theta)$),

$$g(\varphi) \equiv Tr[\hat{\rho}P_{\varphi_0 + \varphi}] = \frac{\nu}{\pi} \int_D d^2\mu(z) p_{\varphi_0 + \varphi}(z)\langle z|\hat{\rho}|z\rangle, \quad p_{\varphi_0 + \varphi}(z) = \frac{1}{2\pi} \frac{1 - r^2}{1 - 2r\cos(\theta - \varphi_0 - \varphi) + r^2}, \tag{5.8}$$

We propose to call $\langle z|\hat{\rho}|z\rangle$ *R-function*, by referring to the name of the regular phase states $|z\rangle$, given by (4.1), which are one-mode coherent states associated to a Holstein-Primakoff realization of the su(1,1) Lie algebra. By writing out the explicite form of the weight in $d^2\mu(z)$ and the integration limits in polar coordinates, we have

$$g(\varphi) = Tr[\hat{\rho}P_{\varphi_0 + \varphi}] = \frac{\nu}{2\pi} \int_0^1 d(r^2) \int_0^{2\pi} d\theta \frac{p_{\varphi_0 + \varphi}(z)}{(1 - r^2)^2}\langle z|\hat{\rho}|z\rangle \quad (0 < \varphi \le 2\pi). \tag{5.8a}$$

It can be shown by a direct calculation that the definition (5.8) of $g(\varphi)$ is consistent with the form of the matrix elements $\langle r|P_{\varphi_0 + \varphi}|s\rangle$ of $P_{\varphi_0 + \varphi}$, as given by (3.13b), in Fock representation. This means, at the same time, that, according to (3.14), in the limit $\nu \to 0$ (provided this exists) the phase density function $g(\varphi)$ reduces to the well-known Loudon phase density,

$$\lim_{\nu \to 0} g(\varphi) = Tr[\hat{\rho}|\varphi_0 + \varphi\rangle\langle\varphi_0 + \varphi|] \quad (0 < \varphi \le 2\pi). \tag{5.8b}$$

As the simplest example for an application of (5.8a), let us calculate the phase probability density function of a number state $|n\rangle$. Since $|\langle z|n\rangle|^2$ does not depend on $\theta$, the angular integral of the Poisson kernel alone is unity for any values of the radial variable $r < 1$. Thus we have

$$g_{|n\rangle}(\varphi) = \frac{\nu}{2\pi} \int_0^1 d(r^2) \int_0^{2\pi} d\theta \frac{p(z)}{(1 - r^2)^2}(1 - r^2)^{\nu+1} \frac{(r^2)^n}{\nu B(\nu, n+1)} = \frac{1}{2\pi}, \tag{5.9}$$

which is a uniform distribution. In calculating the radial integral we have used the definition of the Beta function (B.3c) in Appendix B. This result can be immediately obtained from the first equation of (5.8a), defining equation (5.2) of $P_{\varphi_0 + \varphi}$. For a regular phase state $g(\varphi) = \langle z|P_{\varphi_0 + \varphi}|z\rangle = p_{\varphi_0 + \varphi}(z)$ (according to (5.8a) and (5.2)), thus, in fact, Fig. 2b shows the plot of this distribution. We just have to replace the axes label "phase parameter $\theta$ [$2\pi$]" to "phase variable $\varphi$ [$2\pi$]", and we choose $\theta = \pi/2$ in $z = \rho\exp(i\theta)$. These simple examples illustrate that our regular phase operator $\Phi$ and $N$ are value complementary, they satisfy the conditions introduced by Busch *et al* (2001) on page 5928 of their paper. Since our primary goal in the present paper has been to build up the formalism itself, we leave the discussion of further (non-trivial) examples for future studies.

## 6. Summary

In the present paper we have given a new description of the quantum phase properties of a harmonic oscillator, which may also represent a quantized mode of the radiation field. The primary new element in our formalism is that, by introducing a new 'polar decomposition' of the quantized amplitude of the oscillator, we have been able to build up a reliable correspondence between the classical Fourier serieses and the operator serieses associated to them. Here the word "reliable" means that all the mathematical objects which have been introduced (e.g. phase operator or operator-valued measures) are defined in terms of strongly convergent expressions on a wide enough domain. The new 'exponential phase operator' $F$ contains an (arbitrary) positive parameter v, whose presence secures the strong convergence. We think, convergence questions are also of crucial practical importance, in defining suitable physical quantities, and reconstructing them from measurements. Concerning the mathematical technique itself, it is very convenient that we need not care about convergence questions in rearranging infinite (operator) serieses, for instance. Moreover, this formalism perfectly suits to our classical intuition, though we have exclusively used the abstract space of square-summable sequences (Hilbert-Fock space). We did not rely on less-known representations, like the 'angular representation' in the Hardy-Lebesgue space, used by Garrison and Wong (1970). Both approaches have advantages and disadvantages. For instance, we have not determined the eigenstates of the regular phase operator, thus, the positive operator-valued measures, introduced in the generalized spectral resolution of the new phase operator, are not orthogonal projectors.

The main results of the present paper can be summarized as follows. In Secion 3, by the correspondence of the new 'exponential phase operator' $F$ with the classical Fourier series of the periodic phase, we have constructed a regular phase operator $\Phi$ as a strongly convergent series. This phase operator is a covariant phase observable, according to the definition used by Lahti and Pellonpää (1999, 2000) and Busch *et al* (2001).

In Section 4 we have determined the eigenstates of $F$, which turned out to be SU(1,1) coherent states with Bargmann index $\frac{1}{2}(v+1)$, in the Holstein-Primakoff realization. This result is essentially equivalent to that of Aharonov *et al* (1973), however, they have not applied these states for building up diagonal representations of phase-related operators. Usually, these states are defined by an exponential expression of the su(1,1) Lie algebra basis elements, acting on the vacuum state, and their 'eigenstate property' is not used out. In term of these special SU(1,1) coherent states we have constructed the diagonal representation of the regular phase operator $\Phi$. We note that this expansion is not possible by using the normalized phase states (Shapiro and Shepard 1991), because these latter ones do not form a complete set. The kernels (quantum phase function, phase density and phase projectors) of the diagonal representations are quantum analogons of the corresponding classical Fourier expressions, they perfectly correspond to our intuitive pictures.

In Section 5 we have derived the generalized spectral resolution of the phase operator $\Phi$, and from the positive operator-valued measures, associated to this resolution, we have derived a new phase probability distribution function and phase probability density function. We have shown that in the $\nu \longrightarrow 0$ limit the latter one reduces to the well-known phase density function due to Loudon (1973), which also comes out from the Pegg and Barnett (1989) density if we let the dimension of the truncated Hilbert space going to infinity. We have also introduced the positive definite R-function, as the diagonal matrix element of the density operator taken in the SU(1,1) coherent state (regular phase state) basis.

We think, the new mathematical objects we have constructed may be applied in representing and reconstructing measured phase-related physical quantities, like the carrier–envelope phase difference of ultrashort light pulses. The formalism, worked out in the present paper may, for example be useful in describing the quantum phase properties of extreme fields, like few–cycle or attosecond light pulses.

## Acknowledgments

It is a great honor for the author to dedicate the present paper to celebrating the "150 years of Margarita and Vladimir Man'ko". The author thanks many interesting and fruitful discussions with Prof. Dr. Wolfgang Schleich on the quantum phase problem. This work has been supported by the Hungarian Scientific Research Foundation OTKA, Grant No. K 104260. Partial support by the ELI-ALPS project is also acknowledged. The ELI-ALPS project (GOP-1.1.1-12/B-2012-0001) is supported by the European Union and co-financed by the European Regional Development Fund.

## Appendix A. Derivation of uniform bounds for $F^k$ and $(F^+)^k$

In the present Appendix we derive the inequalities (3.6) for the functions $f_k(n)$ defined in (3.5),

$$f_k(n) \equiv \frac{(n+k)(n+(k-1))\cdots(n+2)(n+1)}{(n+\nu+k)(n+\nu+(k-1))\cdots(n+\nu+2)(n+\nu+1)} = \frac{\Gamma(n+1+k)}{\Gamma(n+1+\nu+k)} \cdot \frac{\Gamma(n+1+\nu)}{\Gamma(n+1)} . \tag{A.1}$$

Since $f_0(n) = 1$ for all $n \geq 0$, below we shall consider the nontrivial cases when $k \geq 1$. Our aim is to derive a formula for the upper bound of $f_k(n)$, in which we separate the $k-$dependence as a negative power of $k$. By using the Stirling formula (see e.g. Titchmarsh 1939, p.150-51) for any positive $x$,

$$\Gamma(x+1) = \left(\frac{x}{e}\right)^x e^{\frac{\Theta_x}{12x}} \sqrt{2\pi x} \quad (0 < \Theta_x < 1) , \tag{A.2}$$

we have

$$f_k(n) = \frac{(n+\nu)^\nu}{(n+k+\nu)^\nu} \cdot \frac{\left(1+\dfrac{\nu}{n}\right)^n}{\left(1+\dfrac{\nu}{n+k}\right)^{n+k}} \cdot \frac{\sqrt{1+\dfrac{\nu}{n}}}{\sqrt{1+\dfrac{\nu}{n+k}}} \cdot \exp\left[\frac{1}{12}\left(\frac{\Theta_{n+k}}{n+k} + \frac{\Theta_{n+\nu}}{n+\nu} - \frac{\Theta_{n+k+\nu}}{n+k+\nu} - \frac{\Theta_n}{n}\right)\right], \qquad \text{(A.3)}$$

where we assumed $n \geq 1$. By using simple algebraic inequivalities and keeping in mind the monotonicity of the exponential function, it can be shown that

$$\exp\left[\frac{1}{12}\left(\frac{\Theta_{n+k}}{n+k} + \frac{\Theta_{n+\nu}}{n+\nu} - \frac{\Theta_{n+k+\nu}}{n+k+\nu} - \frac{\Theta_n}{n}\right)\right] \leq e^{1/4} \quad (k \geq 1, n \geq 1) \qquad \text{(A.4)}$$

Because $(1+(\nu/n))^n$ is monoton increasing as $n \to \infty$ (and converges to $e^\nu$; $(1+(\nu/n))^n \to e^\nu$), the second factor on the right hand side of (A.3) is smaller than 1 (it converges to 1, from below). The third factor is trivially smaller than the square root of $1+\nu$, thus we may write these two estimates as

$$\frac{(1+(\nu/n))^n}{(1+[\nu/(n+k)])^{n+k}} < 1, \quad \frac{\sqrt{1+(\nu/n)}}{\sqrt{1+[\nu/(n+k)]}} < \sqrt{1+\nu} \quad (k \geq 1, n \geq 1). \qquad \text{(A.5)}$$

Thus, according to (A.4-5), in the case $n,k \geq 1$ we receive the following estimate for $f_k(n)$,

$$f_k(n) < e^{1/4}\sqrt{1+\nu}\,\frac{(n+\nu)^\nu}{(n+k+\nu)^\nu} < e^{1/4}\sqrt{1+\nu}\,\frac{(n+\nu)^\nu}{(k+\nu)^\nu} \quad (k \geq 1, n \geq 1). \qquad \text{(A.6)}$$

In the special case $n = 0$, again by using the Stirling formula in (A.1), we have

$$f_k(0) = \frac{\Gamma(1+k)}{\Gamma(1+\nu+k)}\frac{\Gamma(1+\nu)}{\Gamma(1)} = \Gamma(1+\nu)\frac{1}{(k+\nu)^\nu}\,e^\nu\,\frac{1}{(1+(\nu/k))^k}\exp\left[\frac{1}{12}\left(\frac{\Theta_k}{k} - \frac{\Theta_{k+\nu}}{k+\nu}\right)\right]\sqrt{\frac{k}{k+\nu}},$$

$$f_k(0) < \Gamma(1+\nu)\frac{1}{(k+\nu)^\nu}\,e^\nu\,\frac{1}{(1+(\nu/k))^k}\,e^{1/6} \quad (k \geq 1, n = 0). \qquad \text{(A.7)}$$

We know that $(1+(\nu/k))^k \to e^\nu$ as $k \to \infty$, thus it stays finite, and $1 < (1+(\nu/k))^k$ is trivially valid for all $k \geq 1$. Taking this into account, and the second inequality in (A.6), we summarize our results, which are valid for any $\nu > 0$ parameter values,

$$f_k(0) < \Gamma(1+\nu)e^{\nu+1/6}\frac{1}{(k+\nu)^\nu} \quad (k \geq 1, n = 0), \quad f_k(n) < e^{1/4}\sqrt{1+\nu}\,\frac{(n+\nu)^\nu}{(k+\nu)^\nu} \quad (k \geq 1, n \geq 1), \; (\nu > 0). \text{ (A.8)}$$

If $\nu \geq 1$, then in the Stirling formula (A.2) we have $\Theta_\nu/\nu \leq \Theta_\nu < 1$, and in this case the two estimates in (A.8) can be written in a single formula,

$$f_k(n) < b_\nu \frac{(n+\nu)^\nu}{(k+\nu)^\nu}, \quad b_\nu = e^{1/4}\sqrt{2\pi\nu}, \quad (k \geq 1, n \geq 0, \nu \geq 1).$$  (A.9)

For a generic state $|\psi\rangle = \sum_{n=0}^{\infty} c_n |n\rangle$, according to (3.4), (3.5), (2.6) and (2.7), we have

$$F^k|\psi\rangle = \sum_{n=0}^{\infty} c_{n+k} f_k^{1/2}(n)|n\rangle, \quad (F^+)^k|\psi\rangle = \sum_{n=0}^{\infty} c_n f_k^{1/2}(n)|n+k\rangle.$$  (A.10)

The squared norm of the transformed states are

$$\left\| F^k|\psi\rangle \right\|^2 = \sum_{n=0}^{\infty} f_k(n)|c_{k+n}|^2, \quad \left\| (F^+)^k|\psi\rangle \right\|^2 = \sum_{n=0}^{\infty} f_k(n)|c_n|^2.$$  (A.11)

By using the estimates summarized in (A.8), we can derive upper bounds for these norms,

$$\left\| F^k|\psi\rangle \right\|^2 = \sum_{n=0}^{\infty} f_k(n)|c_{k+n}|^2 < \frac{1}{(k+\nu)^\nu}\left[ \Gamma(1+\nu)e^{\nu+1/6}|c_k|^2 + e^{1/4}\sqrt{1+\nu}\sum_{n=1}^{\infty}(n+\nu)^\nu|c_{k+n}|^2 \right],$$  (A.12a)

$$\left\| (F^+)^k|\psi\rangle \right\|^2 = \sum_{n=0}^{\infty} f_k(n)|c_n|^2 < \frac{1}{(k+\nu)^\nu}\left[ \Gamma(1+\nu)e^{\nu+1/6}|c_0|^2 + e^{1/4}\sqrt{1+\nu}\sum_{n=1}^{\infty}(n+\nu)^\nu|c_n|^2 \right].$$  (A.12b)

We can write for the infinite sum on the right hand side of (A.12a)

$$\sum_{n=1}^{\infty}(n+\nu)^\nu|c_{n+k}|^2 \leq \sum_{n=1}^{\infty}(n+k+\nu)^\nu|c_{n+k}|^2 = \sum_{r=k+1}^{\infty}(r+\nu)^\nu|c_r|^2 \leq \sum_{r=1}^{\infty}(r+\nu)^\nu|c_r|^2.$$  (A.13)

Moreover, both $|c_k|^2$ and $|c_0|^2$ are smaller than unity, on the right hand side of (A.12a) and (A.12b), respectively. Accordingly, by taking (A.13) also into account, we can write a uniform bound for the two norms

$$\left\| F^k|\psi\rangle \right\|^2 < \frac{\bar{b}_{\nu,\psi}}{(k+\nu)^\nu}, \quad \left\| (F^+)^k|\psi\rangle \right\|^2 < \frac{\bar{b}_{\nu,\psi}}{(k+\nu)^\nu} \quad (\nu > 0, k \geq 1),$$  (A.14a)

$$\bar{b}_{\nu,\psi} \equiv e^{\nu+1/6}\Gamma(1+\nu) + e^{1/4}\sqrt{1+\nu}\left\langle (N+\nu)^\nu \right\rangle_\psi, \quad \left\langle (N+\nu)^\nu \right\rangle_\psi = \sum_{n=0}^{\infty}(n+\nu)^\nu|c_n|^2 < \infty.$$  (A.14b)

It is important to note that the bound parameter $\bar{b}_{\nu,\psi}$ introduced in (A.14b) does not depend on power index $k$. The bounds in (A.14a) exist and non-trivial for those states for which the $\nu$ − th moments of $(N+\nu)$, introduced in (A.14b) are finite. We denote the set of such states by $\mathscr{D}_\nu(N)$,

$$\mathscr{D}_\nu(N) = \left\{ |\psi\rangle = \sum_{n=0}^{\infty} c_n |n\rangle \in H; \ \sum_{n=0}^{\infty}(n+\nu)^\nu|c_n|^2 < \infty \right\}.$$  (A.15)

It is clear that for $\nu-$values in the range $0 < \nu \le 2$, the domain $\mathscr{D}(N)$ is contained by $\mathscr{D}_\nu(N)$, i.e. $\mathscr{D}(N) \subseteq \mathscr{D}_\nu(N)$, on the other hand, if $\nu \ge 2$, then $\mathscr{D}_\nu(N) \subseteq \mathscr{D}(N)$.

## Appendix B. The regular phase states as SU(1,1) coherent states in Holstein-Primakoff realization

The solution of the eigenvalue equation $F|z\rangle = z|z\rangle$ is straighforward if we expand $|z\rangle$ in a series of Fock states; $|z\rangle = \sum_{n=0}^{\infty} w_n|n\rangle$. By using the definition of $F$ and $E$ given by equations (3.1) and (2.3), respectively, we derive a recurrence relation to the unknown coefficients $w_n$,

$$F|z\rangle = \sum_{k=0}^{\infty} w_k F|k\rangle = \sum_{k=0}^{\infty} |k\rangle w_{k+1} \sqrt{\frac{k+1}{k+1+\nu}} = z\sum_{n=0}^{\infty} w_n|n\rangle, \quad w_{n+1}\sqrt{\frac{n+1}{n+1+\nu}} = zw_n. \tag{B.1a}$$

The solution can be immediately written down

$$w_n = z^n \sqrt{\frac{(\nu+1)(\nu+2)\cdots(\nu+n)}{1 \cdot 2 \cdots n}} w_0 = \sqrt{\frac{(\nu+1)_n}{n!}} z^n w_0, \tag{B.1b}$$

where we have introduced the Pochammer symbol: $(a)_n \equiv \Gamma(a+n)/\Gamma(a) = a(a+1)\cdots(a+n-1)$. We prescribe unit norm $\||z\rangle\| \equiv \sqrt{\langle z|z\rangle} = 1$ for the eigenstate $|z\rangle$, which determines the modulus of the unknown $w_0$. By using the formula for the negative binomial series (Gradshteyn and Ryzhik 1980)

$$(1-x)^{-(\nu+1)} = 1 + (\nu+1)x + \frac{(\nu+1)(\nu+2)}{2!}x^2 + \ldots + \frac{(\nu+1)(\nu+2)\cdots(\nu+k)}{k!}x^k + \ldots, \tag{B.1c}$$

we have

$$\langle z|z\rangle = |w_0|^2 \sum_{n=0}^{\infty} \frac{(\nu+1)_n}{n!}(|z|^2)^n = |w_0|^2 (1-|z|^2)^{-(\nu+1)} = 1, \quad w_0 = (1-|z|^2)^{+\frac{1}{2}(\nu+1)}. \tag{B.1d}$$

Thus, the normalized eigenstates of $F$ can be brought to the equivalent forms

$$F|z\rangle = z|z\rangle; \quad |z\rangle = (1-|z|^2)^{\frac{1}{2}(\nu+1)} \sum_{n=0}^{\infty} \sqrt{\frac{(\nu+1)_n}{n!}} z^n|n\rangle, \quad z \in D = \{z, |z| < 1\}, \tag{B.2a}$$

$$F|z\rangle = z|z\rangle; \quad |z\rangle = (1-|z|^2)^{\frac{1}{2}(\nu+1)} \sum_{n=0}^{\infty} \left[\frac{\Gamma(\nu+1+n)}{\Gamma(\nu+1)n!}\right]^{1/2} z^n|n\rangle, \quad z \in D = \{z, |z| < 1\}. \tag{B.2b}$$

In (B.2b) we have written the expansion coefficients of $|z\rangle$ in the usual form of that of the SU(1,1) coherent states in the discrete series representation (Perelomov 1986). Below we shall use the following parametrization

$$|z\rangle = \sum_{n=0}^{\infty} w_n|n\rangle,$$

$$w_n = (1-\rho^2)^{(\nu+1)/2} c_n \rho^n e^{in\theta} , \quad c_n = \sqrt{\frac{\Gamma(\nu+1+n)}{\Gamma(\nu+1)(n!)}} , \quad z = \rho e^{i\theta} , \quad 0 \le \rho < 1 , \quad 0 \le \theta < 2\pi . \quad \text{(B.2c)}$$

It is interesting to note that the function $f_k(n)$, introduced in (3.5) (or see (A.1)) can be expressed in terms of the factors $c_n$ in (B.2c) as $f_k^{1/2}(n) = c_n / c_{n+k}$.

The states $|z\rangle$ with different $z-$s are not orthogonal, but they form an (over)complete set in the original Hilbert space $H$. This completeness property has been shown by several authors, here we refer the reader to the recent thorough analysis by Vourdas and Wünsche (1998), and references therein. By introducing the suitable weight function, and using the polar coordinates $\theta$ and $\rho$ in the open unit disc $D$, we shall prove

$$\frac{\nu}{\pi} \int_D d^2\mu(z) |z\rangle\langle z| = 1 \equiv \sum_{n=0}^{\infty} |n\rangle\langle n| , \quad d^2\mu(z) = d\rho\rho(1-\rho^2)^{-2} d\theta , \quad z = \rho e^{i\theta} \quad (\nu > 0) . \quad \text{(B.3a)}$$

Having performed the integration with respect to $\theta$, and introducing the integration variable $t \equiv \rho^2$, we obtain the diagonal sum of the diads $|n\rangle\langle n|$

$$\frac{\nu}{\pi} \int_D d^2\mu(z) |z\rangle\langle z| = \nu \int_0^1 dt (1-t)^{\nu-1} \sum_{n=0}^{\infty} |n\rangle c_n^2 t^n \langle n| . \quad \text{(B.3b)}$$

The numerical coefficients $c_n^2$ (see definition in (B.2c) can be alternatively expressed in terms of the Beta function $B(x,y)$ (Euler's integral of the first kind)

$$c_n^2 = \frac{\Gamma(\nu+n+1)}{\Gamma(\nu)\Gamma(n+1)} = \frac{1}{\nu B(\nu, n+1)} , \quad B(x,y) \equiv \int_0^1 t^{x-1}(1-t)^{y-1} dt = \frac{\Gamma(x)\Gamma(y)}{\Gamma(x+y)} \quad (x>0, y>0) . \quad \text{(B.3c)}$$

It can be immediately seen that the radial integral in (B.3b) yields just $B(\nu, n+1)$, whose product with $\nu c_n^2$ equals to unity, thus the completeness relation in (B.3a) has been proved. It is important to emphasize that the $t$-integral in (B.3b) diverges for $\nu = 0$, thus the "normalizable phase states" $|z\rangle_0$ given by (4.6) do not form a complete set.

The states $|z\rangle$ represent the scaled version of the states $|\alpha, k\rangle$ derived already by Aharonov *et al* (1973), as eigenstates of the „generalized destruction operator" $A(k) \equiv \tilde{F}(k)$, where $k$ is a real parameter. This $\tilde{F}(k)$ can be expressed in terms of our $F$ (defined in (3.1)), and an eigenvalue equation analogous to that in (4.1) is also satisfied,

$$F = \tilde{F}(k)\sqrt{\frac{N+k}{N+\nu}} \frac{1}{\sqrt{k+1}} , \quad F = \tilde{F}(k)\frac{1}{\sqrt{k+1}} \quad (\nu = k) , \quad \tilde{F}(k)|z\rangle = \sqrt{k+1} F|z\rangle = \alpha|z\rangle . \quad \text{(B.3d)}$$

The eigenstates $\left|\alpha,k\right\rangle$ given by equation (3.28) in the paper by Aharonov *et al* (1973) can be expressed in term of the states $\left|z\right\rangle$ as

$$\left|z=\frac{\alpha}{\sqrt{k+1}}\right\rangle =$$

$$=\left|\alpha,k\right\rangle =\left[1-\frac{|\alpha|^2}{k+1}\right]^{\frac{1}{2}(k+1)}\sum_{n=0}^{\infty}\left[\frac{\Gamma(\nu+1+n)}{\Gamma(\nu+1)(n!)}\right]^{1/2}\left(\frac{\alpha}{\sqrt{k+1}}\right)^{n}\left|n\right\rangle, \qquad \alpha\in D_\alpha=\{\alpha, |\alpha|<\sqrt{k+1}\}. \qquad \text{(B.3e)}$$

Since the spectrum of $F$ consist of complex numbers in the unit disc $D=\{z, |z|<1\}$, the spectrum of $\widetilde{F}(k)$ covers $|\alpha|<\sqrt{k+1}$. We note that Aharonov *et al* (1973) have also proved that, by keeping $\alpha$ fixed, in the limit $k\to\infty$ the states $\left|\alpha,k\right\rangle$ converge in norm to the well-known coherent states $\left|\alpha\right\rangle$ (Glauber 1963, Sudarshan 1963),

$$\lim_{k\to\infty}\||\alpha,k\rangle-|\alpha\rangle\|=0, \qquad |\alpha\rangle=e^{-\frac{1}{2}|\alpha|^2}\sum_{n=0}^{\infty}\frac{\alpha^n}{\sqrt{(n!)}}|n\rangle. \qquad \text{(B.3f)}$$

This limit behaviour has also been recovered by Gerry and Grobe (1997) in a more general frame, in which they discussed intelligent states associated with the Holstein-Primakoff realization for different values of $\kappa$.

We have seen in Section 4 that the eigenstates in (4.1) are in fact SU(1,1) coherent states in a Holstein-Primakoff-type realization, associated to the Lie algebra of basis operators $K_\pm$ and $K_0$. The definitions and the basic commutators are

$$K_- = A\sqrt{N+\nu}=F(N+\nu), \quad K_+=(K_-)^+=\sqrt{N+\nu}A^+=(N+\nu)F^+, \quad K_+K_-=(N+\nu)N, \qquad \text{(B.4a)}$$

$$K_0=N+\kappa=N+\tfrac{1}{2}(\nu+1), \quad \nu=2\kappa-1, \quad [K_-,K_+]=2(K_0+\kappa), \quad [K_0,K_\pm]\equiv\pm K_\pm. \qquad \text{(B.4b)}$$

By introducing the Hermitian combinations $K_{1,2}$, we have

$$K_1\equiv\tfrac{1}{2}(K_++K_-), \quad K_2\equiv\tfrac{1}{2i}(K_+-K_-), \quad K_1^2+K_2^2=\tfrac{1}{2}(K_-K_++K_+K_-)=(N+\kappa)^2+\kappa-\kappa^2. \qquad \text{(B.4c)}$$

The Casimir operator $C$ takes on the value

$$C\equiv K_0^2-K_1^2-K_2^2=\kappa(\kappa-1), \qquad \kappa=\tfrac{1}{2}(\nu+1)>\tfrac{1}{2}, \qquad \text{(B.4d)}$$

where $\kappa$ is the Bargmann index. The coherent states introduced by Barut and Girardello (1971) are eigenstates of $K_-=F(N+\nu)$, so they differ from $\left|z\right\rangle$, of course.

The proof of (4.7), (4.7a) and (4.8) in Section 4 can be carried out as follows. In order to express the regular phase operator (3.7) in a diagonal form, in terms of the regular phase states (4.1), (B.2b), we use (3.11), the eigenvalue equations $F|z\rangle = z|z\rangle$ and $\langle z|F^+ = z^*\langle z|$, and the completeness relation (4.4), (B.3a),

$$\Phi = \varphi_0 + \pi + \frac{v}{\pi} \int_D d^2\mu(z)(-i)[\log(1 - Fe^{-i\varphi_0})|z\rangle\langle z| - |z\rangle\langle z|\log(1 - F^+ e^{+i\varphi_0})]$$

$$= \varphi_0 + \pi + \frac{v}{\pi} \int_D d^2\mu(z)|z\rangle(-i)\log\left[\frac{1 - ze^{-i\varphi_0}}{1 - z^* e^{+i\varphi_0}}\right]\langle z|$$

(B.5a)

The argument of the logarithm can be brought to the form

$$\frac{1 - ze^{-i\varphi_0}}{1 - z^* e^{+i\varphi_0}} = \frac{1 - \rho\cos(\theta - \varphi_0) - i\rho\sin(\theta - \varphi_0)}{1 - \rho\cos(\theta - \varphi_0) + i\rho\sin(\theta - \varphi_0)} = \frac{1 - ix}{1 + ix}, \text{ where } x \equiv \frac{\rho\sin(\theta - \varphi_0)}{1 - \rho\cos(\theta - \varphi_0)}.$$

(B.5b)

By using the formula 1.622.3 of Gradshteyn and Ryzhik (1980)

$$\frac{1}{2i}\log\left(\frac{1 + ix}{1 - ix}\right) = \arctan x, \text{ i.e. } \frac{1}{i}\log\left(\frac{1 - ix}{1 + ix}\right) = -2\arctan x,$$

(B.5c)

we have

$$-i\log\left(\frac{1 - ze^{-i\varphi_0}}{1 - z^* e^{+i\varphi_0}}\right) = -2\arctan\left[\frac{\rho\sin(\theta - \varphi_0)}{1 - \rho\cos(\theta - \varphi_0)}\right].$$

(B.5d)

By taking the second equation of (B.5a) and (B.5d) into account, we have

$$\varphi_0 + \pi - i\log\left(\frac{1 - ze^{-i\varphi_0}}{1 - z^* e^{+i\varphi_0}}\right) = -i\log\left[e^{i(\varphi_0 + \pi)}\left(\frac{1 - ze^{-i\varphi_0}}{1 - z^* e^{+i\varphi_0}}\right)\right] = -i\log\left[e^{i\varphi(z)}\right] = \varphi(z)$$

(B.6a)

$$e^{i\varphi(z)} = e^{i(\varphi_0 + \pi)}\frac{1 - ze^{-i\varphi_0}}{1 - z^* e^{+i\varphi_0}},$$

(B.6b)

$$\varphi(z) \equiv \varphi(z, z^*) = \varphi_0 + \pi - 2\arctan\left[\frac{\rho\sin(\theta - \varphi_0)}{1 - \rho\cos(\theta - \varphi_0)}\right] \quad (\varphi_0 < \varphi(z) < \varphi_0 + 2\pi).$$

(B.6c)

The lower and upper bounds of $\varphi(z)$ can be derived by taking into account the inequality $-\pi < (-2\arctan x) < \pi$, valid for any real $x$ in the interval $(-\infty < x < \infty)$. From (3.11) one can also show that the diagonal matrix elements of $\Phi$ between the regular phase states just coincide with the *quantum phase function* $\varphi(z)$ defined in equation (B.6c),

$$\langle z|\Phi|z\rangle = \varphi(z) = \varphi_0 + \pi - 2\sum_{k=1}^{\infty} \rho^k \frac{\sin[k(\theta - \varphi_0)]}{k}.$$

(B.7)

The two forms of $\varphi(z)$ in (B.6c) and (B.7) are equivalent, which can be shown by using the Fourier series of the *arctan* function (see e.g. Gradshteyn and Ryzhik 1980, formula 1.448). Since the Fourier series $\sum_{k=1}^{\infty} (\sin k\varphi)/k$ converges to $\pi - \varphi$ in the interval $(0 < \varphi < 2\pi)$, in the limit $\rho \rightarrow 1-0$ the quantum phase function $\varphi(z)$ approaches the classical phase function $\Phi_{cl}(e^{i\varphi})$, given by (2.11) (where we have taken the reference phase $\varphi_0 = 0$).